# Mobility and Energy Conscious Clustering Protocol for Wireless Networks

Abhinav Singh[*], Awadhesh Kumar Singh

Department of Computer Engineering, National Institute of Technology, Kurukshetra, Haryana, India
`abhinavsingh282@gmail.com,aksingh@nitkkr.ac.in`

*Abstract. In this paper we present a distributed clustering protocol for mobile wireless sensor networks. A large majority of research in clustering and routing algorithms for WSNs assume a static network and hence are rendered inefficient in cases of highly mobile sensor networks, which is an aspect addressed here. MECP is an energy efficient, mobility aware protocol and utilizes information about movement of sensor nodes and residual energy as attributes in network formation. It also provides a mechanism for fault tolerance to decrease packet data loss in case of cluster head failures.*

## 1    Introduction

Wireless sensor networks (WSNs) consist of a large number of sensor nodes that are densely deployed in a region of interest and connected through wireless links to collect data about a target or event, and cater a variety of sensing and monitoring applications [1]. In many applications that belong to marine environments, wildlife tracking and protection, and various other such activities, the mobile sensors are more effective as compared to their static counterparts. However, the sensors being energy constrained nodes, the mobile sensors are more prone to crash due to battery exhaustion as mobility incurs more computation overhead and thus it is energy intensive. Hence, the traditional WSN protocols [2, 3] are not suitable for deployment in the environment where sensors are mobile. Therefore, the mobility aware protocols are more preferred option.

In the literature, a number of approaches have been proposed to handle mobility in wireless scenario [8, 9]. Clustering is a popular approach to handle mobility and improve scalability in distributed computing systems [10, 11]. In clustering, the network is partitioned into non overlapped regions   and activities of nodes belonging to each cluster is coordinated by a distinct node called a Cluster Head(CH).Though, CH is responsible for efficient communication and data dissemination, its failure may lead to discontinuity of application. Therefore, for long running applications, fault tolerance is an utmost desirable feature. Secondly, the movement of cluster head may also impact the application in a similar way. Therefore, in this paper, we propose an energy efficient distributed clustering protocol that is fault tolerant and also handles mobility in WSNs.

## 2  System Model

The WSN consists of a set of N nodes connected with wireless links. The nodes in the communication range of a node $N_i$ are called neighbours of $N_i$. The nodes are assumed to be mobile and hence trigger the topological changes. It is also assumed that each sensor node is capable of sensing its velocity, e.g. via an accelerometer or any other sensing hardware embedded in it.

## 3  Related Work

A large number of clustering protocols for sensor nodes have been proposed in the literature. However, due to limitation of space, some representative protocols are being reviewed in the following paragraphs.

The LEACH-Mobile protocol proposed by Kim Do-Seong and Yeong-Jee Chung [5] supports sensor nodes mobility in WSN by adding membership declaration to LEACH protocol. The idea behind this membership declaration is to confirm the inclusion of sensor nodes in a specific cluster during the steady state phase. The CH sends 'data request' message to its members, and receives the data sent back from them. The SN with minimum mobility is elected as cluster head. The LEACH-Mobile outperforms LEACH in terms of packet loss in mobility centric environment. However, it is not traffic and mobility adaptive protocol and there is higher energy waste in idle listening and overhearing of this protocol. [6]

The HEED protocol [4] is a distributed clustering protocol for long-lived ad hoc sensor network in which the main parameter for Cluster Head selection is residual energy levels and leads to a prolonged life of nodes. However it is designed primarily for static sensors and has high data packet drop rates, in case, the nodes are mobile. Therefore, we have proposed a mobile energy conscious protocol (called MECP, henceforth) which is an enhanced version of HEED and it can handle mobility in a more effective way. Moreover, unlike HEED, it is fault tolerant too.

## 4  The MECP concept

### 4.1  Clustering Assumptions

In our protocol, each node takes decisions on the basis of two parameters, namely residual energy and relative velocity with respect to its neighbors. The intra-cluster communication costs are also considered for the clustering process in order to increase the efficiency of energy consumption. For example, cost can be a function of distance from CH or node density of cluster. Generally, a node has several transmission power levels, where the higher power levels can cover greater distances for transmission. We reserve the lower power levels for intra-cluster communication between a normal node and a CH to reduce communication costs. Further, we reserve the higher transmission power levels for inter-cluster communication. Because inter cluster communication with low power levels may lead to link failure or may render unidirectional links. If the power level used for intra-cluster communication is fixed then the cost of communication for a node can be determined by the node degree (D). For instance, to create dense clusters, the cost of communication with a CH is set proportional to 1/D so that dense clusters have lower cost of communication. Alternatively, for load balanced clustering, the cost of communication is set proportional to D so that dense clusters have a higher cost of communication. Consequently, a new node would prefer to join a cluster where the cost of communication is lower. In case, multiple power levels are allowed for intra-cluster communication, we define a cost factor (communication cost factor) CCF where

$$CCF = \frac{\sum_{i=1}^{m} minPower_i}{m} \quad (1)$$

Where, $minPower_i$ is the minimum required power level for communication between a node and its $i^{th}$ neighbor. We also define a velocity factor ($VF$) as

$$VF = \begin{cases} \frac{1}{Va}, & Va \geq 1 \\ 1, & Va < 1 \end{cases} \quad (2)$$

Va is the average of relative velocities between node and its neighbors defined as.

$$Va = \frac{\sum_{i=1}^{m} Relative\ velocity\ of\ Node \wedge neighbour\ i}{m} \quad (3)$$

### 4.2 The working of MECP

The protocol is executed at every node and requires X number of iterations denoted by $X_{it}$. The probability of a node becoming a cluster head is denoted by $CH_{prob}$

$$CH\ prob = K \times \frac{Eres}{Emax} \times VF \quad (4)$$

Where, $K$ is the percentage of nodes that become cluster heads (eg, 10%) initially. *Eres* is the estimated amount of energy remaining, and *Emax*, the maximum energy stored by the battery. The $CH_{prob}$ is restricted in the range [*Pmin*, 1] to allow efficient termination of protocol, will be explained soon. The protocol introduces fault tolerance by allowing each CH to select an Assistant CH(*ACH*) from within its cluster. In case of CH failure, due to reasons like physical damage, depletion of energy or movement of CH out of communication range, etc., the cluster members suffer data packet loss. After timeout, the ACH assumes the role of CH. In such case, the cluster members resend the data packet to the new CH. The new CH possesses all the updated routing information that the previous CH had. Therefore, it successfully avoids any further loss of data packet and renders application continuity.

The initialization phase starts with initializing the values of $L_{adj}$ using neighbor discovery(Fig. 1). Afterwards, each node broadcasts its cost and velocity to its neighbors and calculates its $CH_{prob}$(Fig.2).

In the second phase, every node searches for a CH or a tentative CH and in the absence of one, the nodes become a tentative CH with a probability $CH_{prob}$. At the end of each iteration, the value of $CH_{prob}$ doubles and the process continues till the value of $CH_{prob}$ becomes $\geq 1$. In the third phase, a node, either becomes a Final CH and broadcasts *Declare_FinalCH* message to its neighbours or it becomes a member of a cluster(Fig. 3 and Fig. 4).

In the last phase, each CH selects its ACH and broadcasts the ACH ID to its member nodes (Fig. 5). In case of failure of CH during protocol operation, ACH takes charge of CH. In case, a node moves out of its cluster during protocol operation, it requests to join another cluster in its transmission range.

The protocol is carried out as explained by the pseudocode below.

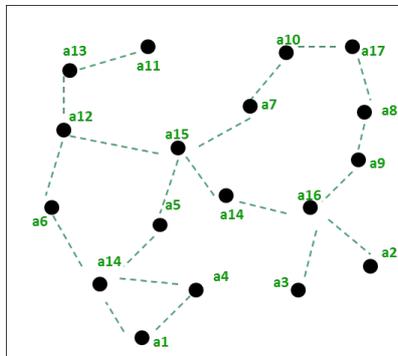

Fig. 1 Neighbor Discovery

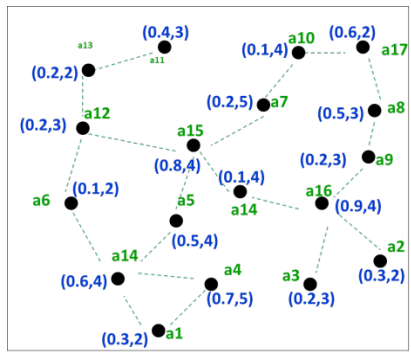

Fig. 2 Compute $CH_{prob}$ and cost

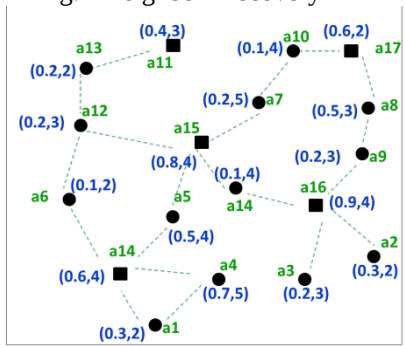

Fig. 3 Cluster Head Election

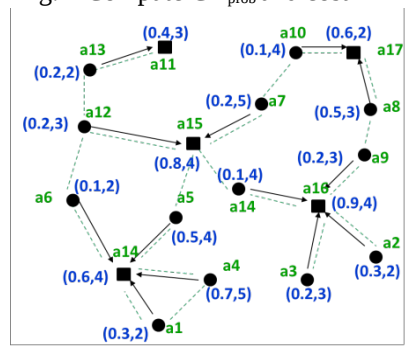

Fig. 4 Normal nodes select their CH

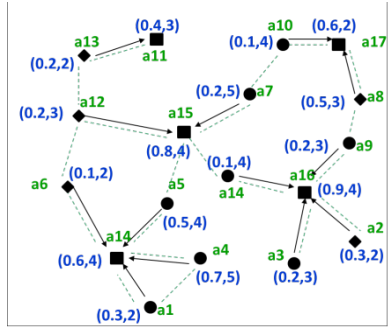

Fig. 5 Selection of ACH by each CH

### 4.3 Pseudo Code

**Phase I    Initialization**
1) $L_{adj}$ ← Add all nodes in communication range to list
2) Compute Cost of communication for each node in $L_{adj}$ and broadcast cost, my_velocity
3) $\text{CH prob} = K \times \dfrac{Eres}{Emax} \times VF$       //Initialise value of $CH_{prob}$
4) Set final_CH = False
5) Set Assistant_CH = False

**Phase II**
   Repeat
1) $L_{CH}$ ← All nodes in $L_{adj}$ which are final CH or Tentative CH
2) If $L_{CH} \neq \phi$               // List is not empty
   {     my_CH = least_cost($L_{CH}$)     //select least cost node as CH
         If(my_ch = =node_ID)
         {    If($CH_{prob}$ = 1)
              {    Declare_FinalCH(NodeID,Cost)
                   Set final_CH ← True}
              Else
              {    Declare_Tentative_CH(NodeID,Cost) }
         }
   }
   ElseIf ($CH_{prob}$ = = 1)
   {    Set my_CH = nodeID
        Declare_FinalCH(NodeID,Cost)
    Set Final_CH = true}
    Elseif (Random (0,1) ≤ $CH_{prob}$)
    {    Declare_Tentative_CH(NodeID,Cost)}

3) $CH_{prev}$ = $CH_{prob}$
4) $CH_{prob}$ = min ($CH_{prob}$ *2, 1)
5) Until $CH_{prev}$ = 1

**Phase III        Termination**
   1) If (final_CH = = False )

```
            {   L_FCH ← All Final Cluster Heads in L_CH
            1a)     If(L_FCH ≠ φ)
                {   My_CH = least_cost(L_FCH)
                    Declare_join(Cluster_head_ID,Node_ID,cost) }
        1b)     Else
                {   Set Final_CH = True
                    L_Members ← All nodes that have elected to join
                    my cluster.
                    Declare_FinalCH(NodeID,Cost) }
            }

        2) Else

            {
                Set Final_CH = True
                L_Members ← All nodes that have elected to join my
cluster.
                Declare_FinalCH(NodeID,Cost)
            }

        proc recieve_message (recieve declaration of formation of
final CH or tentative CH)
        {   Lch<-- save  nodeID along with identifier final or
            tentative }

        After completion of clustering,
        Proc select_ACH(NodeID,cost)
        {       if(node is a CH)
                {L_Members ← All nodes that have elected to join my
                        cluster.
                My_ach= least_cost(L_Members )
                Declare_ACH(NodeID of ACH,cost);}

                Else
                {proc  recieve_message  (recieve  declaration  of
                formation of final CH or tentative CH)}
        }
    Phase IV During T_p
        If(packet data is lost when sending data to CH)
        {
  1            Resend data to ACH
  2            If(packet data is lost when sending data to
               ACH)
  3            {
  4            L_FCH ← All CHs in neighbourhood
  5            My_CH= least_cost(L_FCH)
  6            Declare_join(Cluster_head_ID,Node_ID,cost)
  7            My_ACH ← Receive ID of ACH node from CH
  8            }

        }
        proc recieve_message (declaration of new ACH by CH )
        {   my_ACH = NodeID of ACH recieved;}
```

### 4.4 Proof of Correctness

**Lemma i].** MECP terminates in $X_{iter}$ = O(1) iterations
**Proof.** The worst case scenario is when a node has very low residual energy and very high Va. In this case, the $CH_{prob}$ is equal to $P_{min}$. Now since the $CH_{prob}$ doubles every iteration and terminates when it finally reaches a value $\geq 1$, therefore we have

$$Pmin * 2^{Xiter-1} \geq 1 \quad \text{and}$$
$$Xiter \leq \lceil \log_2 \frac{1}{Pmin} \rceil + 1$$

Therefore $X_{iter} \approx O(1)$.

Since essentially this translates into $\leq \lceil \log_2 \frac{1}{CHprob} \rceil + 1$, a node with higher $E_{res}$ and lower Va will terminate its MECP execution much faster than other nodes and will hence this will allow low energy or highly mobile nodes to join its cluster.

**Lemma ii].** Any node that wishes to join the WSN, will do so by the end of MECP.
**Proof.** Let us assume that the node isn't a part of the WSN and hence isn't a CH or a regular node by the termination of MECP. This means line 1 of Phase III is satisfied while 1a isn't, this means that 1b shall execute and the node becomes a CH, which is a contradiction.

**Lemma iii].** No node can be a part of more than one cluster by the end of MECP.
**Proof.** Let us assume a node is part of two clusters. This means that line 1 and 1a of phase III must have been executed, after which the node becomes part of one cluster and end MECP execution. There is no provision for it to execute lines 1 and 1a again in the same cycle, which contradicts our assumption.

## 5 Inter-cluster communication

The CHs aggregate data over a round and send the aggregated data in that round in single transmission. However, in certain situations, the need of inter-cluster communication may arise. The inter-cluster communication in MECP is multi-hop through various CHs. Nevertheless, in the mid of inter-cluster communication, a CH may move beyond the range of transmission of other CHs incurring packet data loss because it doesn't reach the sink and the data for the entire round from a particular cluster can be lost. Though, an appropriate recovery mechanism may help in recovery of lost data, it can also be avoided by using gateway nodes that are also called guard node as proposed in DEMC [7] protocol. Guard nodes are intermediate nodes that help in transmission of data, in case, two CHs are not within transmission range of each other.

## 6  Conclusion

We presented a multiphase distributed clustering protocol that is energy efficient and also effectively handles mobility of nodes during application execution. Out of the three phases of clustering, only second phase may involve multiple iterations nevertheless that are bounded by parameter *Pmin*. This feature makes the process of reclustering lightweight in addition to significant reduction in the latency involved in clustering. Furthermore, the clustering related decisions in MECP are based primarily on local information and therefore MECP suffers limited message overhead.